\documentclass[apj]{emulateapj}
\usepackage{apjfonts}

\shorttitle{ACTIVITY CYCLES OF SOUTHERN ASTEROSEISMIC TARGETS}
\shortauthors{METCALFE ET AL.}

\begin{document}
 
\title{Activity Cycles of Southern Asteroseismic Targets}

\author{T.~S.\ Metcalfe, P.~G.\ Judge}
\affil{High Altitude Observatory, NCAR, P.O.\ Box 3000, Boulder, CO 80307}

\author{S.\ Basu}
\affil{Department of Astronomy, Yale University, P.O.\ Box 208101, New Haven, CT 06520}

\author{T.~J.\ Henry}
\affil{Department of Physics and Astronomy, Georgia State University, Atlanta, GA 30302}

\author{D.~R.\ Soderblom} 
\affil{Space Telescope Science Institute, 3700 San Martin Dr., Baltimore MD 21218}

\author{M.\ Kn{\"o}lker, M.\ Rempel}
\affil{High Altitude Observatory, NCAR, P.O.\ Box 3000, Boulder, CO 80307}

\slugcomment{Proceedings of Solar Analogs II}

\begin{abstract}

The Mount Wilson Ca HK survey revealed magnetic activity variations in a 
large sample of solar-type stars with timescales ranging from 2.5 to 25 
years. This broad range of cycle periods is thought to reflect differences 
in the rotational properties and the depths of the surface convection 
zones for stars with various masses and ages. Asteroseismic data will soon 
provide direct measurements of these quantities for individual stars, but 
many of the most promising targets are in the southern sky (e.g., 
$\alpha$~Cen~A~\&~B, $\beta$~Hyi, $\mu$~Ara, $\tau$~Cet, $\nu$~Ind), while 
long-term magnetic activity cycle surveys are largely confined to the 
north. In 2007 we began using the SMARTS 1.5-m telescope to conduct a 
long-term monitoring campaign of Ca~{\sc ii}~H~\&~K emission for a sample 
of 57 southern solar-type stars to measure their magnetic activity cycles 
and their rotational properties when possible. This sample includes the 
most likely southern asteroseismic targets to be observed by the Stellar 
Oscillations Network Group (SONG), currently scheduled to begin operations 
in 2012. We present selected results from the first two years of the 
survey, and from the longer time baseline sampled by a single-epoch survey 
conducted in 1992.

\end{abstract}



\section{ASTROPHYSICAL CONTEXT}

Astronomers have been making telescopic observations of sunspots since the 
time of Galileo, gradually building a historical record showing a periodic 
rise and fall in the number of sunspots every 11 years. We now know that 
sunspots are regions with an enhanced local magnetic field, so this 
11-year cycle actually traces a variation in surface magnetism. Attempts 
to understand this behavior theoretically often invoke a combination of 
differential rotation, convection, and meridional flow to modulate the 
field through a magnetic dynamo \citep[e.g., see][]{rem06,dg06}. 
\cite{wil78} was the first to demonstrate that many solar-type stars 
exhibit long-term cyclic variations in their Ca~{\sc ii}~H~\&~K emission, 
analogous to the solar variations.

Significant progress in dynamo modeling could only occur after 
helioseismology provided meaningful constraints on the Sun's interior 
structure and dynamics \citep{bro89,sch98}. Variations in the mean 
strength of the solar magnetic field lead to significant shifts 
($\sim$0.5~$\mu$Hz) in the frequencies of even the lowest-degree p-modes 
\citep{lw90,sal04}. Space-based asteroseismology missions, such as MOST 
\citep{wal03}, CoRoT \citep{bag06}, and Kepler \citep{bor07}, as well as 
future ground-based networks like the Stellar Oscillations Network Group 
(SONG), will soon allow additional tests of dynamo models using other 
solar-type stars \citep[see][]{cha07,met07}. Ironically, many of the best 
targets for asteroseismology are stars in the southern sky, while 
long-term activity cycle surveys have largely been confined to the north.

\section{SURVEY METHODOLOGY}

In August 2007, we began a long-term Ca HK monitoring program for a sample 
of 57 southern solar-type stars using the 1.5-m telescope at CTIO. This 
survey will continue through July 2010 as an NOAO long-term observing 
program, supplemented by additional time from our collaborators at SMARTS 
institutions. By the end of the long-term program, we will either directly 
measure or provide firm lower limits on the cycle period for those stars 
with the shortest activity cycles. By comparing our observations with 
those from an earlier single-epoch survey \citep{hen96}, we will also 
establish interesting limits on the cycle period for those stars with the 
slowest variations in activity.

The single-epoch survey of \cite{hen96} contained a total of 1016 
observations of 815 individual stars with visual magnitudes between 0.0 
and about 9.0, which were observed using the {\it RC Spec} instrument on 
the CTIO 1.5-m telescope. Several sub-samples were defined, including the 
``Best \& Brightest'' (B) and ``Nearby'' (N) samples, which together 
contain 92 individual stars with visual magnitudes between 0.0 and 7.9, 
and B$-$V colors that are approximately solar. We further restrict our 
sample to the 57 stars in the combined (B+N) sample that are brighter than 
V$=6$, the limiting magnitude of future ground-based asteroseismic 
observations by SONG (see Table~\ref{tab1}). All of the most promising 
southern asteroseismic targets ($\alpha$~Cen~A~\&~B, $\beta$~Hyi, 
$\mu$~Ara, $\tau$~Cet, 70~Oph~A, $\nu$~Ind) are included in this B+N 
subsample.

The CTIO 1.5-m telescope is now operated by a consortium of about a dozen 
partners, known as SMARTS (Small and Moderate Aperture Research Telescope 
System). This consortium runs the telescope in queue mode, with 
observations collected by a trained technician and made available for 
download by the principal investigator. The technician cycles between the 
available instruments based on the demand for each during the semester. It 
is important to note that SMARTS operates the {\it only} southern 
telescope run in queue mode with an aperture and instrument that are 
appropriate for this project. Aside from a dedicated survey telescope like 
the one at Mount Wilson, SMARTS is the only option that makes such 
time-domain monitoring feasible.

\section{INITIAL RESULTS}

\begin{figure*}[t]
\centerline{\includegraphics[angle=0,width=\linewidth]{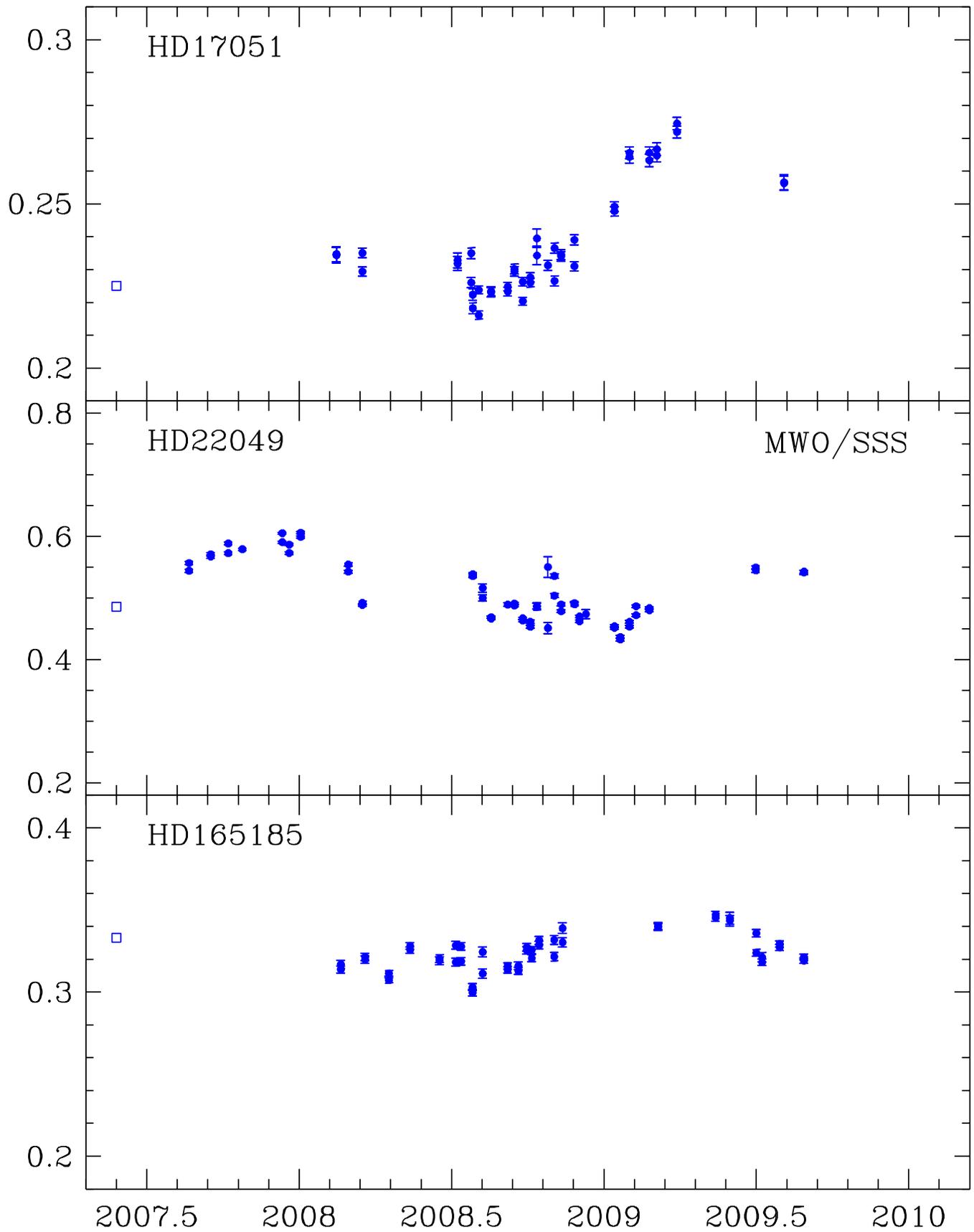}}
\caption{Early highlights from our activity cycle survey, showing the 
Mount Wilson S-index for several of our program stars during the first two 
years. The open square near the left axis shows the mean activity 
level measured by \cite{hen96} for comparison.\label{fig1}}
\end{figure*}

The S-index derived from our first two years of survey data are shown in 
Figure~\ref{fig1} for several stars with interesting variations over this 
short time baseline. The southern target HD 17051 ($\iota$~Hor) exhibits a 
steady rise from mid-2008 through early 2009, with the most recent data 
suggesting a possible reversal from peak activity in mid-2009. The active 
equatorial target HD 22049 ($\varepsilon$~Eri) shows smooth variations 
from one season to the next, with a mean activity level comparable to that 
measured by the Mount Wilson and Lowell surveys. The moderately southern 
G0V star HD 165185 appears to have a maximum in activity in early 2009. 
The mean activity level measured by \cite{hen96} is shown for each star as 
an open square near the left axis for reference. No longer-term trends are 
apparent for these three targets.

\begin{figure*}[t]
\centerline{\includegraphics[angle=0,width=4.5in]{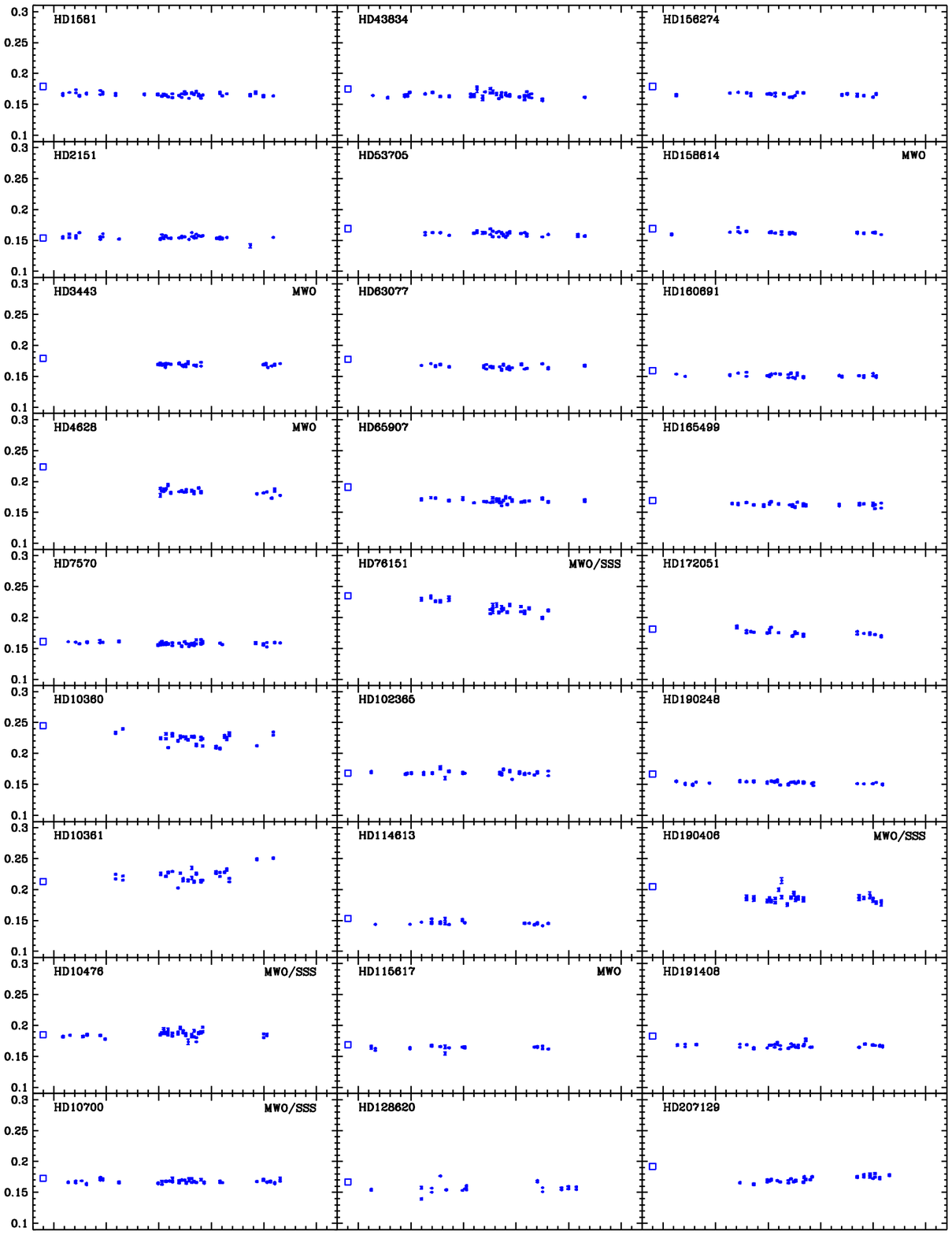}}
\vskip -0.105in
\centerline{\includegraphics[angle=0,width=4.5in]{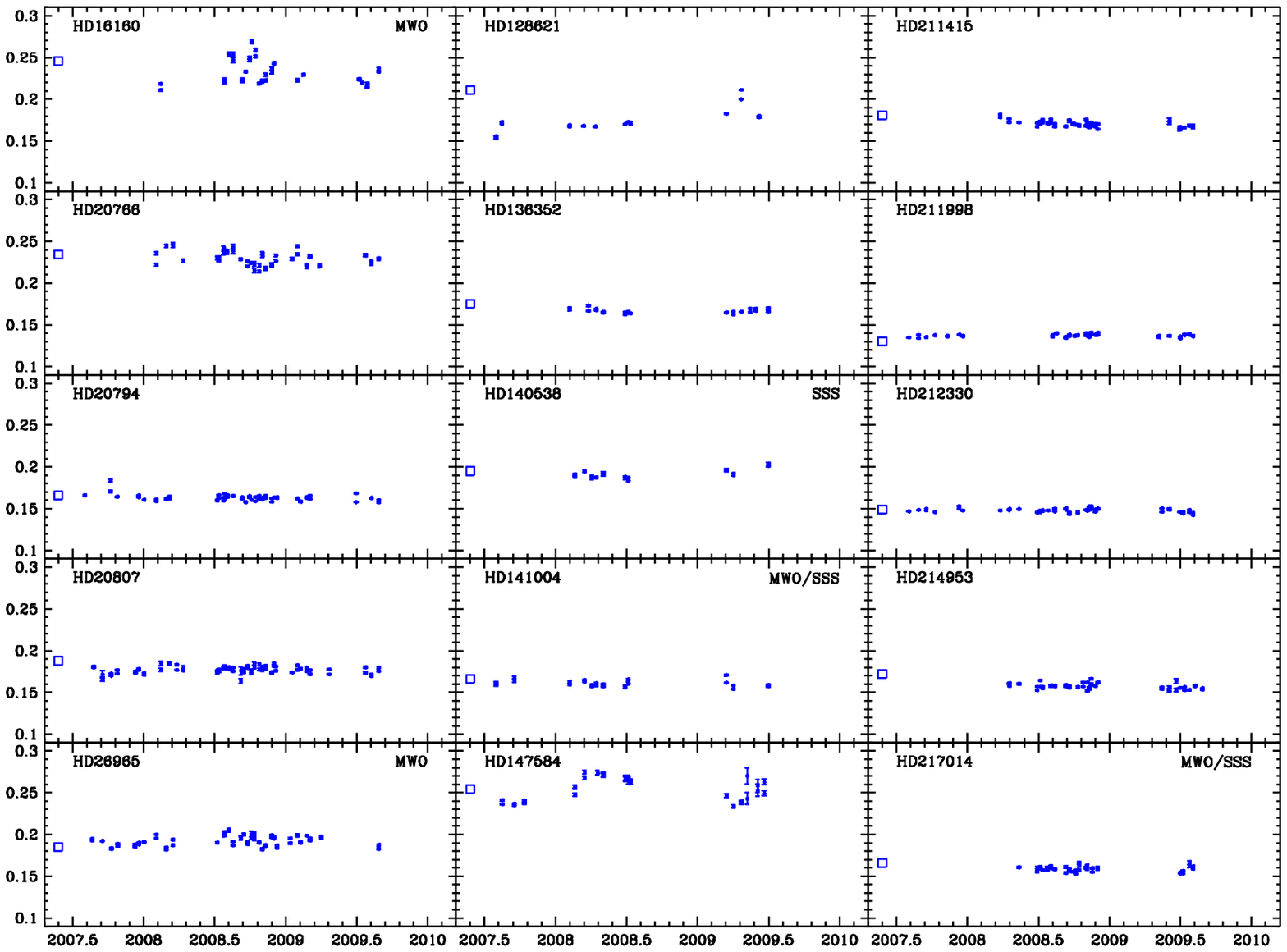}}
\caption{The first two years of data for low activity stars in our 
survey.\label{fig2}}
\end{figure*}

\begin{figure*}[t]
\centerline{\includegraphics[angle=0,width=4.5in]{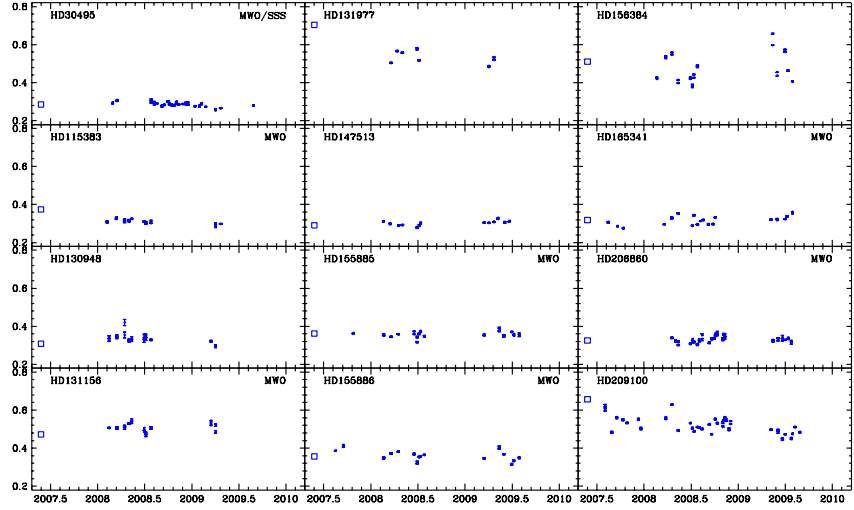}}
\caption{The first two years of data for high activity stars in our 
survey.\label{fig3}}
\end{figure*}

Similar plots for all of our survey targets are shown in Figures 
\ref{fig2} \& \ref{fig3}. The sample has been divided into subsets of low- 
and high-activity stars, each with a common vertical scale. Within each 
subset, the stars are displayed in three columns ordered by their HD 
number. Activity levels from the single-epoch survey of \cite{hen96} are 
again shown on the left to reveal longer-term trends, and targets that are 
included in the Mount Wilson (MWO) and Lowell (SSS) surveys are indicated 
at the upper right of each panel.

We performed least-squares linear fits to each of the data sets to 
quantify the significance of any increasing or decreasing trends over the 
first two years, with the results shown in Table~\ref{tab1}. We found 
significant slopes (from 4.5 to 126$\sigma$) for 85\% of the targets, 
while the remaining 15\% were statistically flat (from 1.4 to 
3.8$\sigma$). This is roughly consistent with the fraction found to be in 
a Maunder Minimum type phase by \cite{hen96}. Significant trends are 
observed for the recent asteroseismic targets $\beta$~Hyi (HD~2151, 
7.3$\sigma$), $\tau$~Cet (HD~10700, 8.4$\sigma$), $\alpha$~Cen~B 
(HD~128621, 116$\sigma$), $\mu$~Ara (HD~160691, 8.8$\sigma$), 70~Oph~A 
(HD~165341, 64$\sigma$), and $\nu$~Ind (HD~211998, 6.3$\sigma$). No 
significant trend is yet seen for $\alpha$~Cen~A (HD~128620, 1.7$\sigma$).

\acknowledgments
This survey began with SMARTS time purchased by High Altitude Observatory 
through Georgia State University, and is currently supported under NOAO 
long-term program 2008B-0039. The National Center for Atmospheric Research 
is a federally funded research and development center sponsored by the 
U.S.~National Science Foundation.


\begin{table*}
\caption{Characteristics of the 57 stars in our survey sample.\label{tab1}}
\renewcommand{\arraystretch}{1.2}
\hskip 1.0in
\begin{tabular}[t]{|lccccc|lccccc|}
\hline
HD    &  RA(2000)  &  Dec(2000)  &   V   & B$-$V & Trend? &
HD    &  RA(2000)  &  Dec(2000)  &   V   & B$-$V & Trend? \\
\hline
   1581    & 00~20~04.3 & $-$64~52~29 & 4.226 & 0.576 & $\searrow$    & 
 131156~w  & 14~51~23.4 & $+$19~06~02 & 4.700 & 0.730 & $\searrow$    \\ 
   2151    & 00~25~45.1 & $-$77~15~15 & 2.820 & 0.618 & $\searrow$    & 
 131977    & 14~57~28.0 & $-$21~24~56 & 5.723 & 1.024 & $\searrow$    \\ 
   3443~w  & 00~37~20.7 & $-$24~46~02 & 5.572 & 0.715 & $\searrow$    & 
 136352    & 15~21~48.1 & $-$48~19~03 & 5.652 & 0.639 & $\rightarrow$ \\ 
   4628~w  & 00~48~23.0 & $+$05~16~50 & 5.742 & 0.890 & $\searrow$    & 
 140538~s  & 15~44~01.8 & $+$02~30~55 & 5.865 & 0.684 & $\nearrow$    \\ 
   7570    & 01~15~11.1 & $-$45~31~54 & 4.959 & 0.571 & $\searrow$    & 
 141004~ws & 15~46~26.6 & $+$07~21~11 & 4.422 & 0.604 & $\rightarrow$ \\ 
  10360    & 01~39~47.7 & $-$56~11~34 & 5.900 & 0.800 & $\searrow$    & 
 147513    & 16~24~01.3 & $-$39~11~35 & 5.385 & 0.625 & $\nearrow$    \\
  10361    & 01~39~47.2 & $-$56~11~44 & 5.800 & 0.860 & $\nearrow$    & 
 147584    & 16~28~28.1 & $-$70~05~04 & 4.900 & 0.555 & $\nearrow$    \\ 
  10476~ws & 01~42~29.8 & $+$20~16~07 & 5.242 & 0.836 & $\nearrow$    & 
 155885~w  & 17~15~21.0 & $-$26~36~10 & 5.330 & 0.860 & $\rightarrow$ \\ 
  10700~ws & 01~44~04.1 & $-$15~56~15 & 3.495 & 0.727 & $\searrow$    & 
 155886~w  & 17~15~20.8 & $-$26~36~09 & 5.290 & 0.850 & $\searrow$    \\ 
  16160~w  & 02~36~04.9 & $+$06~53~13 & 5.791 & 0.918 & $\searrow$    & 
 156274    & 17~19~03.8 & $-$46~38~10 & 5.330 & 0.770 & $\searrow$    \\ 
  17051    & 02~42~33.5 & $-$50~48~01 & 5.400 & 0.561 & $\nearrow$    & 
 156384    & 17~18~57.2 & $-$34~59~23 & 5.910 & 1.040 & $\nearrow$    \\ 
  20766    & 03~17~46.2 & $-$62~34~31 & 5.529 & 0.641 & $\searrow$    & 
 158614~w  & 17~30~23.8 & $-$01~03~47 & 5.314 & 0.715 & $\rightarrow$ \\ 
  20794    & 03~19~55.7 & $-$43~04~11 & 4.260 & 0.711 & $\searrow$    & 
 160691    & 17~44~08.7 & $-$51~50~03 & 5.127 & 0.694 & $\searrow$    \\
  20807    & 03~18~12.8 & $-$62~30~23 & 5.239 & 0.600 & $\nearrow$    & 
 165185    & 18~06~23.7 & $-$36~01~11 & 5.900 & 0.615 & $\nearrow$    \\ 
  22049~ws & 03~32~55.8 & $-$09~27~30 & 3.726 & 0.881 & $\searrow$    & 
 165341~w  & 18~05~27.3 & $+$02~30~00 & 4.026 & 0.860 & $\nearrow$    \\
  26965~w  & 04~15~16.3 & $-$07~39~10 & 4.426 & 0.820 & $\nearrow$    & 
 165499    & 18~10~26.2 & $-$62~00~08 & 5.473 & 0.592 & $\searrow$    \\ 
  30495~ws & 04~47~36.3 & $-$16~56~04 & 5.491 & 0.632 & $\searrow$    & 
 172051    & 18~38~53.4 & $-$21~03~07 & 5.858 & 0.673 & $\searrow$    \\ 
  43834    & 06~10~14.5 & $-$74~45~11 & 5.080 & 0.714 & $\searrow$    & 
 190248    & 20~08~43.6 & $-$66~10~55 & 3.554 & 0.751 & $\searrow$    \\ 
  53705    & 07~03~57.3 & $-$43~36~29 & 5.559 & 0.624 & $\searrow$    & 
 190406~ws & 20~04~06.2 & $+$17~04~13 & 5.788 & 0.600 & $\rightarrow$ \\ 
  63077    & 07~45~35.0 & $-$34~10~21 & 5.363 & 0.589 & $\searrow$    & 
 191408    & 20~11~11.9 & $-$36~06~04 & 5.315 & 0.868 & $\rightarrow$ \\ 
  65907    & 07~57~46.9 & $-$60~18~11 & 5.595 & 0.573 & $\rightarrow$ & 
 206860~w  & 21~44~31.3 & $+$14~46~19 & 5.945 & 0.587 & $\nearrow$    \\ 
  76151~ws & 08~54~17.9 & $-$05~26~04 & 6.000 & 0.661 & $\searrow$    & 
 207129    & 21~48~15.8 & $-$47~18~13 & 5.579 & 0.601 & $\nearrow$    \\ 
 102365    & 11~46~31.1 & $-$40~30~01 & 4.892 & 0.664 & $\searrow$    & 
 209100    & 22~03~21.7 & $-$56~47~10 & 4.688 & 1.056 & $\searrow$    \\ 
 114613    & 13~12~03.2 & $-$37~48~11 & 4.849 & 0.693 & $\rightarrow$ & 
 211415    & 22~18~15.6 & $-$53~37~37 & 5.363 & 0.614 & $\searrow$    \\ 
 115383~w  & 13~16~46.5 & $+$09~25~27 & 5.209 & 0.585 & $\searrow$    & 
 211998    & 22~24~36.9 & $-$72~15~19 & 5.290 & 0.650 & $\nearrow$    \\ 
 115617~w  & 13~18~24.3 & $-$18~18~40 & 4.739 & 0.709 & $\nearrow$    & 
 212330    & 22~24~56.4 & $-$57~47~51 & 5.310 & 0.665 & $\searrow$    \\ 
 128620    & 14~39~36.5 & $-$60~50~02 & 0.010 & 0.710 & $\rightarrow$ & 
 214953    & 22~42~36.9 & $-$47~12~39 & 5.988 & 0.584 & $\searrow$    \\
 128621    & 14~39~35.1 & $-$60~50~14 & 1.350 & 0.900 & $\nearrow$    & 
 217014~ws & 22~57~28.0 & $+$20~46~08 & 5.469 & 0.666 & $\searrow$    \\
 130948    & 14~50~15.8 & $+$23~54~43 & 5.863 & 0.576 & $\searrow$    & 
 $\cdots$  &$\cdots$    &$\cdots$    &$\cdots$&$\cdots$&$\cdots$      \\
\hline
\end{tabular}
\centerline{\footnotesize w/s: Mount Wilson / Lowell target}
\end{table*}

\end{document}